\documentclass[referee]{raa}           
\usepackage{CJK}
\usepackage{graphicx,times}
\usepackage{natbib}
\usepackage{amssymb,amsmath}
\usepackage{lscape}
\usepackage{longtable}
\bibpunct{(}{)}{;}{a}{}{,}

\usepackage[pagebackref=true]{hyperref}

\begin{document}

\title{Magnetic activities and parameters of 43 flare stars in the GWAC archive}
\begin{CJK*}{UTF8}{gbsn}

 \volnopage{ {\bf 20XX} Vol.\ {\bf X} No. {\bf XX}, 000--000}
   \setcounter{page}{1}

\author{Guang-Wei Li (李广伟)\inst{1}, Chao Wu (吴潮)\inst{1}, Gui-Ping Zhou (周桂萍)\inst{2}, Chen Yang (杨晨)\inst{3},
    Hua-Li Li (黎华丽)\inst{1}, Jie Chen (陈洁)\inst{2},Li-Ping Xin (辛立平)\inst{1},Jing Wang (王竞)\inst{4},
    Hasitieer Haerken (哈斯铁尔·哈尔肯)\inst{5,1},Chao-Hong Ma (马超红)\inst{3},Hong-Bo Cai (蔡洪波)\inst{1},
    Xu-Hui Han (韩旭辉)\inst{1},Lei Huang (黄垒)\inst{1},Xiao-Meng Lu (卢晓猛)\inst{1},Jian-Ying Bai (白建迎)\inst{1},
    Xu-Kang Zhang (张旭康)\inst{3},Xin-Li Hao (郝新丽)\inst{3},Xiang-Yu Wang (王祥玉)\inst{6,7},Zi-Gao Dai (戴子高)\inst{6,7},
    En-Wei Liang (梁恩维)\inst{4},Xiao-Feng Meng (孟小峰)\inst{3},Jian-Yan Wei (魏建彦)\inst{1}
   }

   \institute{Key laboratory of Space Astronomy and Technology, National Astronomical Observatories, Chinese Academy of Sciences, Beijing 100101, China; {\it lgw@bao.ac.cn}\\
   \and
             Key laboratory of solar activity, National Astronomical Observatories, Chinese Academy of Sciences, Beijing 100101, China\\
   \and
        Lab of Web and Mobile Data Management, Renmin University of China, Beijing 100872, China\\
   \and 
       Guangxi Key Laboratory for Relativistic Astrophysics, School of Physical Science and Technology, Guangxi University, Nanning 530004, Peopleʼs Republic of China\\
   \and
       School of Artificial Intelligence of Beijing Normal University, No.19, Xinjiekouwai St, Haidian District, Beijing, 100875, China\\
    \and
        School of Astronomy and Space Science, Nanjing University, Nanjing 210093, People's Republic of China \\
    \and
        Key Laboratory of Modern Astronomy and Astrophysics (Nanjing University), Ministry of Education, Nanjing 210093, People's Republic of China\\   
\vs \no
   {\small Received 20XX Month Day; accepted 20XX Month Day}
}

\abstract{In the archive of the Ground Wide Angle Camera (GWAC), we found 43 white light flares from 43 stars, among which, 
three are sympathetic or homologous flares, and one of them also has a quasi-periodic pulsation with a period of $13.0\pm1.5$ 
minutes. Among these 43 flare stars, there are 19 new active stars and 41 stars that have available TESS and/or \textit{K2} light 
curves, from which we found 931 stellar flares. We also obtained rotational or orbital periods of 34 GWAC flare stars,  of which 33 
are less than 5.4 days, and ephemerides of three eclipsing binaries from these light curves. Combining with low resolution spectra 
from LAMOST and the Xinglong 2.16m telescope, we found that $L_{\rm H\alpha}/L_{\rm bol}$ are in the saturation region in the 
rotation-activity diagram. From the LAMOST medium-resolution spectrum, we found that Star \#3 (HAT 178-02667) has double 
H$\alpha$ emissions which imply it is a binary, and two components are both active stars. Thirteen stars have flare frequency 
distributions (FFDs) from TESS and/or \textit{K2} light curves. These FFDs show that the flares detected by GWAC can occur at a 
frequency of 0.5 to 9.5\,yr$^{-1}$. The impact of flares on habitable planets was also studied based on these FFDs, and flares from 
some GWAC flare stars may produce enough energetic flares to destroy 
ozone layers, but none can trigger prebiotic chemistry on their habitable planets.
}

   \authorrunning{G.-W. Li et al. }            
   \titlerunning{Flare Stars in the GWAC archive}  
   \maketitle

%

\section{Introduction} \label{sec:intro}
 \end{CJK*}
Stellar flares are powerful explosions that can occur on stars ranging from A-type
\citep{bal15} to even L-type \citep{2020MNRAS.494.5751P}. 
Solar flares have been well studied based on space observatories, such as the \emph{Geostationary 
Operational Environmental Satellite} (GOES), \emph{Solar Dynamics Observatory} \citep{pes12}, 
the \emph{the Reuven Ramaty High Energy Solar Spectroscopic Imager} \citep{lin02}, and
the \emph{Interface Region Imaging Spectrograph} \citep{de21}, etc. They may release explosive energy via
magnetic reconnection \citep{2009ARA&A..47..291Z} and 
released by electromagnetic radiations from radio to $\gamma$-ray \citep[e.g.][]{bai89, ost05}, 
and coronal mass ejections (CMEs) \citep[e.g.][]{kah92}. The stronger the flare, the more 
likely it is to produce a CME \citep{li21}.
Since the Sun belongs to an ordinary incative star of G2V type (Balona 2015), it is worthy of 
investigating stellar flare activities from most kinds of active stars to understand whether stellar 
flares may experience similar physics to solar flares or not.
 \par
Statistical studies of stellar flares have been conducted based on many survey projects.
In the space, the Kepler \citep{bor10} provided photometric data with high precision, which can be
used to study flare activities on stars across the H-R diagram with homogenous data for 
the first time \citep{bal15,yang19}.  The Transiting Exoplanet Survey Satellite \citep[TESS;][]{ric15}
surveys all the sky, covering much wider than Kepler/K2. As a result, many new flare stars with high photometric
precision can be studied \citep[e.g.][]{tu21,how22}. On the ground, the Next Generation Transit Survey
\citep{whe18}, ASAS-SN \citep{sha14}, Evryscope \citep{law15}, and so on, have achieved 
fruitful results on flare study \citep[e.g.][]{jac21,rod20,how20a}. 

\par
Rotation is a key parameter to decide the activity of a star, e.g., factor in inducing stellar flares. 
Some indicators of stellar activity show the 
well-known activity-rotation relationship with a critical period. For stars with rotation periods 
smaller than the critical period, the activity is saturated, otherwise the activity decreases as 
the rotation period increases. The stellar activity-rotation relationship has been identified by
X-ray \citep{piz03}, white light \citep{rae20}, Ca II H \& K \citep{zhang20} and H$\alpha$ \citep{newton17,yang17,lu19}.
\par
Pre-main sequence stars often show intense flare activities, which result in the hot plasma escaping 
and then angular momentum losses \citep{col19}. Magnetized stellar winds also can brake stellar 
rotations \citep{gal13}. Therefore, with rotation slowing dwon, flare activity decreases with age \citep{ilin21,dav19}.  
The same mechanism is also proposed to occur in a close binary system \citep{yak05}. 
Magnetic braking may result in the shrink of the orbit period, and then make both components in the binary 
synchronously spin up \citep{qian18}, and thus the  more frequent flare activity.
\par
Flare activity may play key roles in affecting habitability of nearby planets in the way of UV irradiation and CMEs. 
For M stars, on one hand, M stars can not produce enough UV photos \citep{rim18}, so UV radiation from frequent flares 
is needed to contribute to the creation of primitive life \citep{xu18};  on the other hand, UV radiation from frequent flares 
can also destroy ozone layers and life would not survive \citep{til19}. Moreover, CMEs from flares can erode 
even the whole atmosphere of a habitable exoplanet \citep{lam07,atri21}.
\par
In this paper, we present 43 stellar white light flares in the archive of the Ground-based Wide Angle Cameras (GWAC). 
GWAC is one of ground facilities of the Space-based multi-band astronomical Variable Objects Monitor
\citep[SVOM;][]{wei18}, in order to detect the optical  transits with a cadence of 15 seconds \citep{xin21,wang20}.
We searched all light curves during December 2018 and May 2019 of stars with Gaia G $<15$ mag, and 
found 43 stellar flares form 43 stars. In Section \ref{sec:data}, we will introduce the light curves we used;
In Section \ref{sec:prof}, we will show three sympathetic or homologous flares and one quasi-periodic pulsation;
In Section \ref{sec:bin},  four binaries are studied; In Section \ref{sec:ha}, we will present the rotation-activity relationship by 
H$\alpha$ emissions;  The impacts of flares on habitable planets are discussed in Section \ref{sec:hab}; At last, Section 
\ref{sec:con} is conclusion.

\section{Light Curves} \label{sec:data}
The GWAC stellar flare candidates between December 2018 and May 2019 were obtained by the program given by \citet{ma19}, 
which tried to find flares by a wavelet algorithm.
We inspected all candidates by eye and found 43 stellar flares from 43 stars. We checked these stars in SIMBAD \footnote{\url{http://simbad.u-strasbg.fr/Simbad}} and the International Variable Star Index 
\footnote{\url{https://www.aavso.org/vsx/}} , and found that 19 stars have never been reported as flare stars or having H$\alpha$ 
emissions, thus new active stars. All GWAC flares are listed in Table \ref{tab:gwacflare}. 
We searched their light curves from the MAST site \footnote{\url{https://mast.stsci.edu}},  and found {\it TESS} and {\it K2} light 
curves for 39 stars. For the stars that have both K2 and TESS light curves, the TESS light curves were used.
For TESS light curves, we noticed that there are several products for the same sector  from different groups, 
and if light curves of the Science Processing Operations Center \citep[SPOC;][]{jen16} are available, 
then use them, otherwise use light curves of TESS-SPOC \citep{cal20}. The light curves of simple aperture 
photometry (SAP) were used, because pre-search data conditioning (PDC) ones may remove real variabilities 
\citep{vida19}. We also checked the PDC light curves of our sample, and found they work as well as SAP ones.
Star \#3 (HAT 178-02667), \#14 (1RXS J075908.2+171957), \#24, and \#38 (BX Ari) have no available light curves in the MAST 
site. Star \# 3 (HAT 178-02667) is not observed by TESS, and Star \#24 is contaminated by a very bright star 13 arcseconds away. 
As a result, we obtained light curves of Star \#14 (1RXS J075908.2+171957) and \#38 (BX Ari) from their Full Frame Images. In 
sum,  41 of 43 GWAC flare stars have TESS or K2 light curves. The TESS sectors and K2 campaigns used in this work are listed 
in Table \ref{tab:sector}.
\par

\subsection{Flare Detection and Rotational Periods}
\begin{figure}
\centering
\includegraphics[width=12.0cm, angle=0]{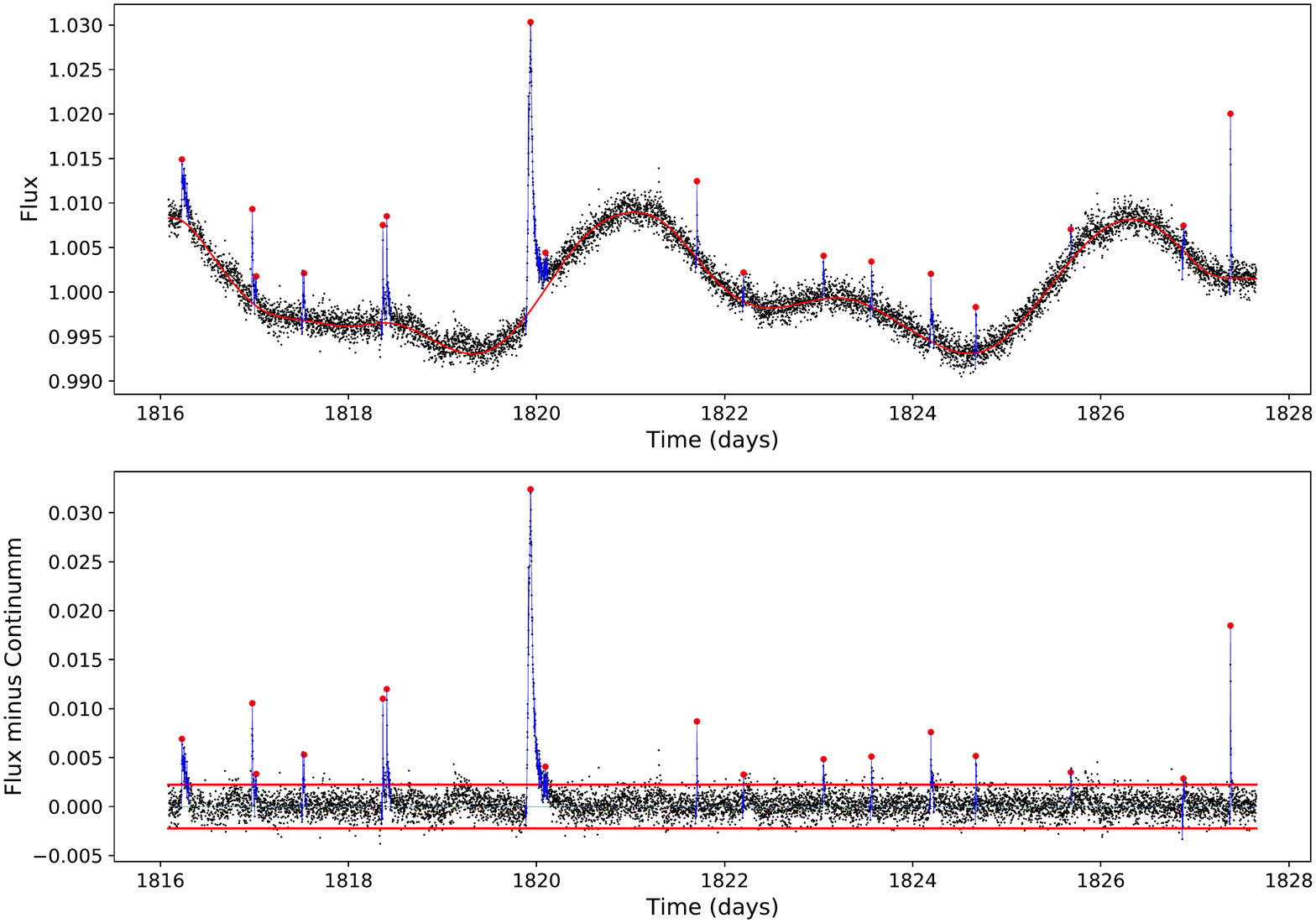}
\caption{ Upper panel: the black dots are from a part of the  light curve of Star \#1 (TIC 141533801) in Sector 19 of TESS. 
The red curve is the fitted cubic B-Spline curve.  Lower panel:  the black dots are the residuals of the fluxes minus the fitted 
cubic B-Spline curve, and the two red lines indicate $\pm 3\sigma$ positions. In both panels, the detected 
flares are shown in blue lines and their peaks are indicted by red dots.
\label{fig:findflare}}
\end{figure}

To detect flares, for each light curve of a TESS sector or K2 campaign, we used a cubic B-spline 
to fit the out-of-flare variability, and then removed the fitted B-spline from the light curve. In the residual of 
a light curve, flares can be detected by residual fluxes larger than 3$\sigma$, where $\sigma$ is the 
standard deviation of the out-of-flare residual. At last, the stellar rotational period was calculated from the
out-of-flare variability. Detailed steps are as follows:
\begin{itemize}
\item Step 1: Remove the points with fluxes greater than the top $2\%$ flux from the light curve ($l_0$), 
the new light curve is denoted as $l_1$.
\item Step 2:  Calculate the Lomb-Scargle periodogram \citep[LS; ][]{lomb76,sca82} of $l_1$ 
using the code  {\tt LombScargle} in the python package {\tt  astropy.timeseries} \citep{astropy13,astropy18}, 
then determine the preriod $P_0$ of the light curve.
\item Step 3: A cubic B-Spline curve with a knot interval of 0.1$P_0$ is used to fit the $l_1$, and denote the 
new B-Spline as $S_1$ and $R_1 = l_1 - S_1$. Calculate the standard deviation $\sigma_1$ of $R_1$, and remove the 
points with $R_1$ values greater than 2$\sigma_1$. The new light curve is denoted as $l_2$, and $l_1 := l_2$. Then repeat this 
step again, and obtain the the standard deviation $\sigma$ of $l_1$, B-Spline $S$, and $R = l_0 - S$ for finding flares.
\item Step 4: A flare is detected from $R$ if there are at least  3 (for curves of 2 minutes cadence) or 2 (for curves of 30 minutes 
cadence) successive fluxes greater than 2$\sigma$ and the flare peak is greater than 3$\sigma$.
\item Step 5: After all flares were removed from the light curve, the rotational period $P$ was calculated from the
out-of-flare light curve by LS.
\end{itemize}
Figure \ref{fig:findflare} shows the results of our algorithm. In the upper  panel, the red line is the fitted cubic B-spline, and flares 
detected are shown in blue. All flares detected were inspected by eye, and finally obtained 931 flares.
\par
The rotational periods of 31 stars were obtained by the above algorithm, and the periods of 3 eclipsing binaries were calculated in 
Section \ref{sec:bin}. Star \#3 (HAT 178-02667) is not observed by TESS and K2, but it has a period of 1.717885 days from \citet{har11}, which 
may be the orbital period (see Section \ref{sec:bin} for details), so there are total 35 GWAC flare stars have periods. Among the 
31 stars that have rotational periods, 30 stars have periods less than 5.4 days, and the left one has a period of about 10.42 days, 
and thus all are rapid rotators.
\par
Flares and periods detected in TESS and K2 light curves by the above algorithm, the 43 GWAC flare light curves and 
the flare movie of Star \#4 (G 176-59), \#7 LSPM J1542+6537), \#28 (G 235-65), and \#39 (1RXS J064358.4+704222) are all given in \url{https://nadc.china-vo.org/res/r101145/}. 
 
 \subsection{Flare Energy}
The equivalent duration (ED) \citep{ger72} was used to calculate a flare energy. ED is defined as:  
$ \sum_{i}\frac{\bigtriangleup f_i}{f}  \times \bigtriangleup t$, where, $\bigtriangleup f_i$ is a flux variation at time $t_i$ in the flare, 
$f$ is the quiescent flux and $\bigtriangleup t$ is the cadence. Then the flare energy is $E = ED \times f$.
\par

GWAC has no filter, and the Gaia $G$ photometry of Gaia DR3 \citep{gaia22} was used to calibrate the GWAC photometry by 
the GWAC pipeline, with a photometric accuracy better than 0.1 mag. Thus, we used the Gaia $G$ band filter and passband zero 
to calculate the quiescent flux $f$ of a GWAC flare star. The Gaia $G$ band filter was from \citet{rie21}, the Vega 
spectrum was from \citet{cas94}, and the Gaia $G$ magnitude of Vega was set 0.03 mag \citep{jor10}. As a result, the passband 
zero $p_0$ of Gaia $G$ band is $9.12 \times 10^{-6}$ erg\,cm$^{-2}$\,s$^{-1}$. 
\par
The distances of GWAC flare stars were calculated from parallaxes of Gaia DR3, but there are three stars 
(Star \#8 (HAT 149-01951), \#14 (HAT 149-01951),  and \#29 (HAT 149-01951)) without available parallaxes in Gaia DR3. Then we 
used the equation of $M_{Ks} = 1.844+1.116(V-J)$ given by \citet{rae20}, and $d = 10^{0.2(M_{Ks}-Ks-5)}$ in pc to calculate their 
distances, where Ks and J are from 2MASS \citep{skr06}, and V is from APASS \citep{hen15}. As a result, the quiescent flux in the 
$G$ band of a star is: $f = \pi d^2 p_0 10^{-0.4G}$ erg\,s$^{-1}$.

\par
To obtain flare energies of TESS flares, the TESS response function and the passband zero from \citet{sul17} were used. 
The observed quiescent flux $f_o$ of a star in the TESS passband was the median value of its light curve multiplied by the TESS 
passband zero. Thus the the quiescent flux is $f = \pi d^2 f_o$.

Two stars: Star \#15 (HAT 307-06930) and \#36 (CU Cnc) only have {\it K2} light curves. Then the formulae (1) - (6) in 
\citet{shi13} were used to calculate quiescent fluxes in the Kepler passband\footnote{The Kepler response function was from \url{https://keplergo.github.io/KeplerScienceWebsite/the-kepler-space-telescope.html}}, with the surface temperatures and stellar radii from \citet{hub16}.
\par
We assumed the fare energy was from a blackbody radiation with a temperature of 9000 K, and then used observed flare 
energies in Gaia $G$, TESS or $K2$ filters to calculate whole white light flare energies, which should be lower than 
the true released flare energies \citep[e.g.][]{haw91,kre11}. 

\subsection{Spectra}
The Guoshoujing Telescope \citep[the Large Sky Area Multi-Object Fiber Spectroscopic Telescope, LAMOST;][]{cui12} can obtain 
4000 spectra in one exposure, and is located at the Xinglong Observatory, the same place as GWAC and the 2.16m telescope. In 
LAMOST DR8  \footnote{\url{http://www.lamost.org/dr8/v1.1/}}, there are about 11 million low-resolution spectra 
(LRS, $R \sim 1800$) and 6 million medium-resolution spectra (MRS, $R \sim 7500$). We searched spectra of GWAC flare stars 
in LAMOST DR8, and obtained available LRS for 25 stars, and MRS for 13 stars. Because 11 stars have available both LAMOST 
LRS and MRS, there are 27 stars have LAMOST spectra. We also obtained LRS of another 7 flare stars  by the 2.16m 
telescope with the instrument G5. The spectral resolution is about 2.34 \AA\,pixel$^{-1}$, and the wavelength coverage is 5200 
\AA - 9000 \AA  \citep{zhao18}. In sum, 32 of 43 flare stars have LSR, and 7 have LAMOST MRS.
\par
The spectroscopic standards from \citet{kir91} were used to assign spectral types of M stars that have LRS from LAMOST or 
the 2.16m telescope. For stars that have spectral types earlier than M0, their spectral types are from LAMOST DR8.
\par
\begin{figure}
\centering
   \includegraphics[width=12.0cm, angle=0]{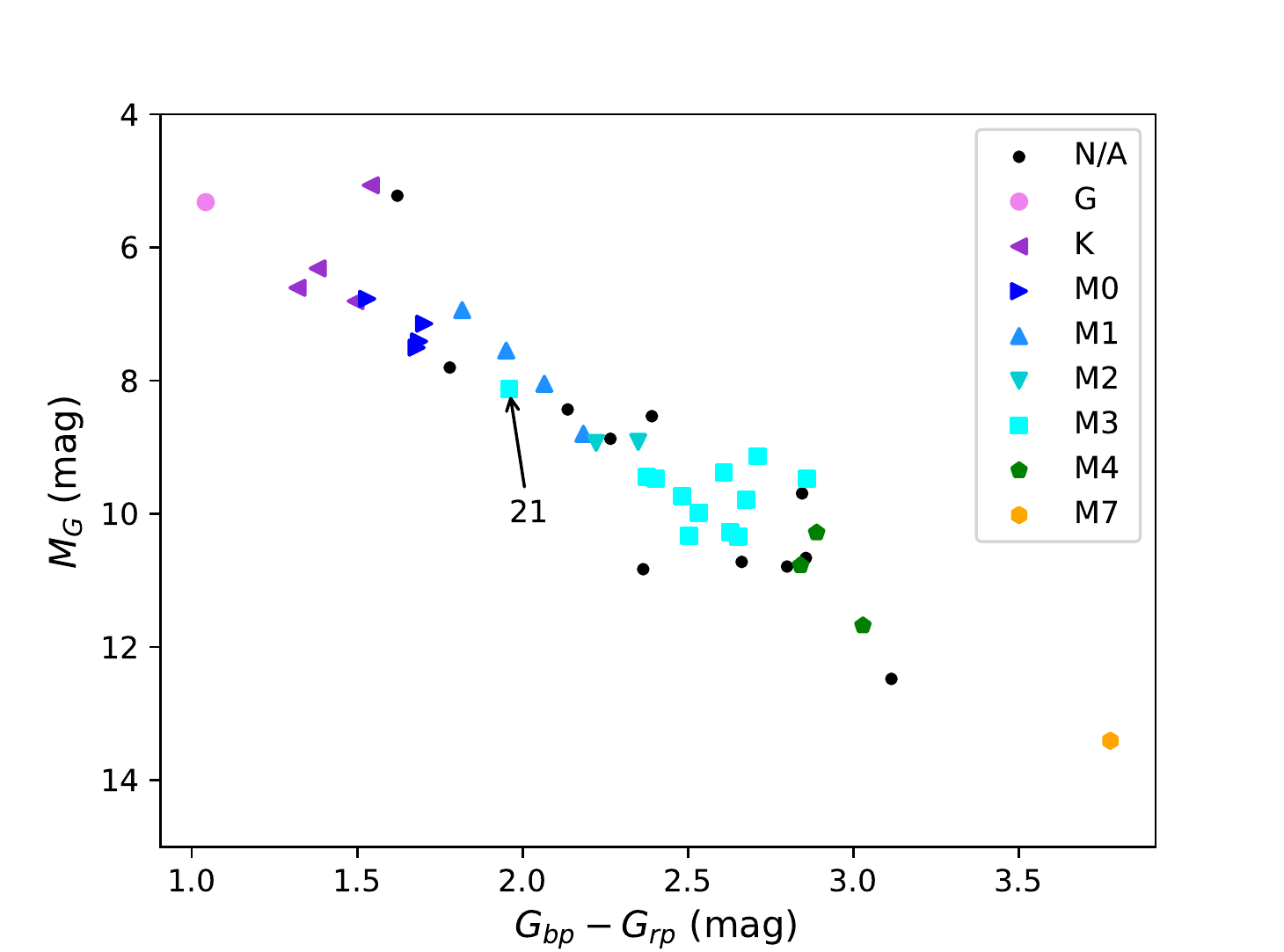}
\caption{The Color-Magnitude Diagram (CMD) of flare stars with their spectral types. $M_G$ is the absolute magnitude of 
Gaia $G$. $G$, $G_{bp}$ and $G_{rp}$ are from Gaia DR3. The 31 stars that have LRS are shown in different symbols and 
colors for different spectral types. The black dots are the stars that have not available  spectra in this paper.
\label{fig:cmd}}
\end{figure}

The Color-Magnitude Diagram (CMD) of flare stars with their spectra types are shown in Figure \ref{fig:cmd}, where Gaia $G$, 
parallax $\varpi$ ( in milliarcsecond), $G_{bp} - G_{rp}$ are from Gaia DR3, and the absolute magnitude 
$M_G  = G + 5 {\rm log}_{10}(\varpi) -10$. 
For 32 stars that have LRS, their H$\alpha$ are all shown in emission, which indicate they 
are all active stars. Among them, thirty-one stars were assigned spectral types and are shown in different symbols and colors in 
Figure \ref{fig:cmd}. Star \#21 was assigned a spectral type of M3, but with a bluer color in Figure \ref{fig:cmd}. We found that its 
$ruwe = 2.478$ in Gaia DR3, which implies that there may be some astrometric problems or it is not a single star, thus the Gaia 
photometry is unreliable. Star \#18 (DR Tau) is not in Figure \ref{fig:cmd}, because it is a T Tauri star \citep{cha79}, and there is no 
available absorption line in its spectrum for spectral classification.
\par
For the other 11 stars without available spectra in this paper, their spectral types were also assigned by their $G_{bp} - G_{rp}$ in 
Figure \ref{fig:cmd}. Among them, Star \#9 \citep[2MASS J04542368+1709534; ][]{her14}, \#23
\citep[1RXS J101627.8-005127; ][]{zic05}, \#27 \citep[1RXS J120656.2+700754; ][]{chr01}, \#28 \citep[G 235-65; ][]{rei04}, \#30 
\citep[1RXS J082204.1+744012; ][]{fle88}, and \#35 \citep[1RXS J075554.8+685514; ][]{zic05} had been identified as active stars 
in the literature.
\par
\par

\begin{figure}

\centering
   \includegraphics[width=12.0cm, angle=0]{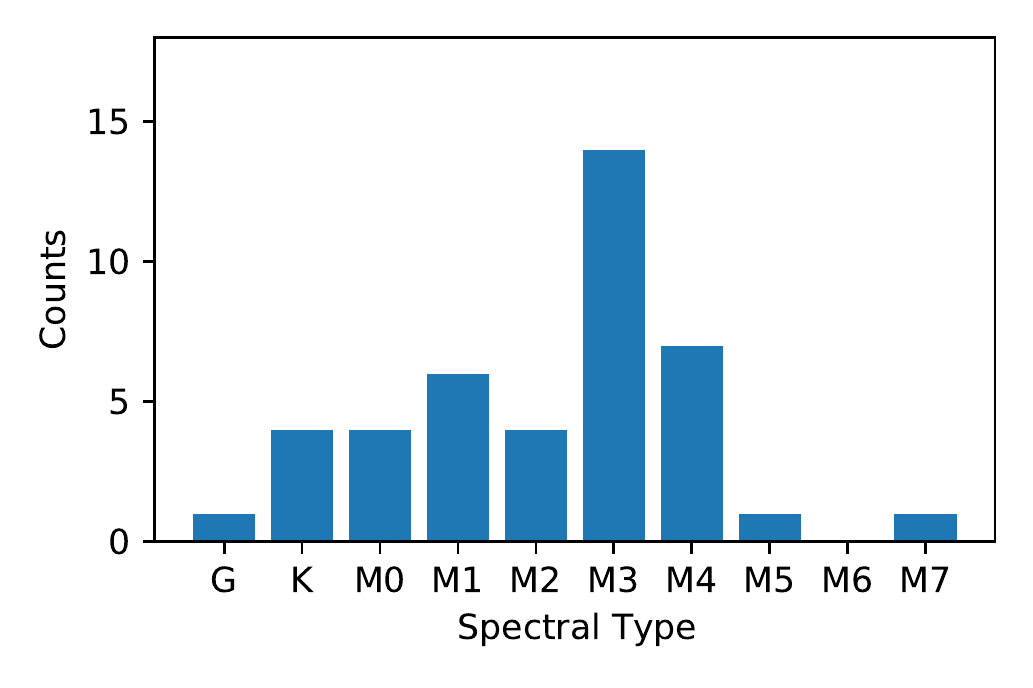}
\caption{The distribution of spectral types of GWAC flare stars, except one T Tauri (Star \#18; DR Tau).
\label{fig:spt}}
\end{figure}

In sum, there are one G type star, four K type stars, thirty-seven M type stars, and one T Tauri in our sample. Figure \ref{fig:spt} 
shows the distribution of spectral types of GWAC flare stars, except one T Tauri (Star \#18; DR Tau). 

\par
To calculate the H$_\alpha$ emission luminosity $L_{\rm H_{\alpha}}$, the S\'ersic function \citep{ser68} was used to fit 
H$_\alpha$ emission profiles. The function is: $F(\lambda) = A_0\exp(Z^{A_3}/A_3) + A_4 + A_5\lambda$, where 
$Z=|(\lambda-A_1)/A_2|$, $\lambda$ is wavelength in \AA, and $A_i$, $i=0, \cdots, 5$ are coefficients to be fitted. The 
wavelength range was set to $6564.61 \pm 20$ \AA (6564.61 \AA is the vacuum  wavelength of H$\alpha$), and the LAMOST 
spectral luminosity of H$\alpha$: $L^{\prime}_{\rm H_{\alpha}} = \int_{6544.61}^{6584.61} A_0\exp(Z^{A_3}/A_3)  d\lambda$. 
The fluxes of LAMOST spectra are not calibrated, but there is only a constant ratio between each spectrum and its true flux. For 
each star that has LAMOST LRS, we calculated its spectral flux $f_L$ from its LRS spectrum in the SDSS r$^{\prime}$ filter 
\citep{fuk96}, and its observed flux $f_A$ from r$^{\prime}$ given by APASS \citep{hen15}. 
Then its  $L_{\rm H\alpha} = L^{\prime}_{\rm H\alpha} * f_A / f_L$.
\par
To obtain the quiescent bolometric luminosity $L_{\rm bol}$ of a star, photometric data from r$^{\prime}$, g$^{\prime}$ and 
i$^{\prime}$ of APASS \citep{hen15},  J, H, and K$_s$ of 2MASS \citep{skr06}, and W1, W2, W3, and W4 of WISE \citep{jar11} 
for each star, were fitted by a blackbody irradiation function. In the calculation, the reference wavelengths and the zero points of 
all filters were from the SVO Filter Profile Service \footnote{\url{http://svo2.cab.inta-csic.es/theory/fps/}} \citep{rodrigo20}. 
Finally, $L_{\rm H\alpha} / L_{\rm bol}$ can be obtained.  $L_{\rm H\alpha} / L_{\rm bol}$ that can be calculated from spectra are listed in Table \ref{tab:halum}. Because one star can have several LAMOST spectra, and one spectra has one $L_{\rm H\alpha} / L_{\rm bol}$, so there are several $L_{\rm H\alpha} / L_{\rm bol}$ for a star in Table \ref{tab:halum}.

\par

\par
The information of GWAC flare stars with their GWAC flares are given in Table \ref{tab:gwacflare}.

\section{Flare Profiles} \label{sec:prof}
\subsection{Sympathetic or homologous flares}
Based on an enormous amount of observations, solar flares are well known to be produced by magnetic reconnections in active 
regions \citep[ARs;][]{tori19}.  Solar ARs are the regions full of intense magnetic fields with complex morphologies 
\citep{mci90,sam00}. Successive flares are often observed on the Sun and are identified as sympathetic activity from different 
regions with physical causal links \citep{pea90,moo02,sch15,hou20}, or homologous ones occurring in the same AR 
\citep[e.g.][]{louis21,xu14}. 
\par
\begin{figure}

\centering
   \includegraphics[width=12.0cm, angle=0]{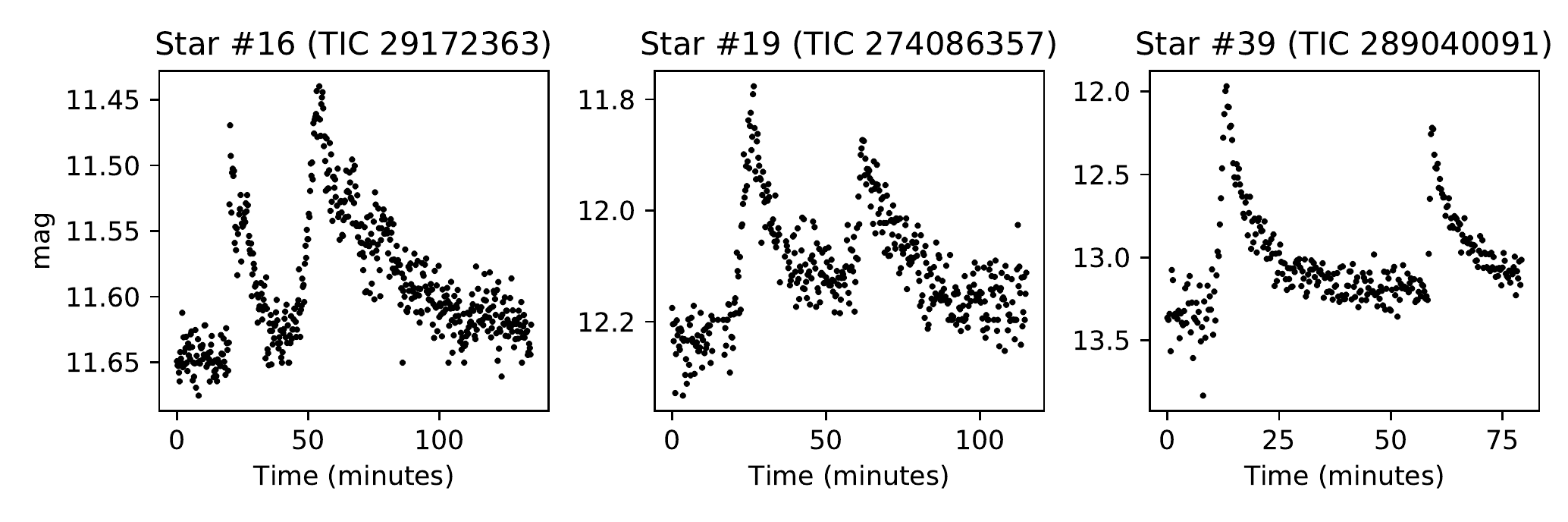}
\caption{Successive flares from Star \#16 (TIC 29172363), \#19 (TIC 274086357; 1RXS J031750.1+010549), and \#39 (TIC 289040091; 1RXS J064358.4+704222).
\label{fig:successive}}
\end{figure}

It is interesting that there are three pairs of flares successively appearing among the 43 GWAC samples as shown in Figure 
\ref{fig:successive}. Each pair of flares occurred following a similar profile of light curves in the interval of 20-50 minutes. The 
morphology of the light curves during a flare may reflect the release processes of magnetic energies, and manifest complex 
magnetic topology of the intense magnetic fields. The successive occurrence of three pairs of stellar flares indicate that the cool 
stars may share the similar physical magnetic explosions to those sympathetic flares or homologous ones happening on the Sun. 
With the development of future imaging observations for the other stars, it is anticipated to unveil these possibly universal 
physical mechanisms.

\par
\subsection{Quasi-periodic pulsation}
Quasi-periodic pulsations (QPPs) are very common phenomenons in solar flares \citep{van16,sim15,kup10}, but QPPs in stellar 
white light flares are still rare \citep{how22,pug16}. More than a dozen of mechanisms were suggested to trigger QPPs 
\citep{kup20}. Coronal loop lengths can be derived from periods of QPPs from  some mechanisms with some theoretical 
considerations \citep{ram21}. The periods of QPPs in white light curves are of tens minutes \citep{ram21,pug16}, but at short 
cadence (20 seconds) of TESS, QPP periods less than 10 minutes were also found \citep{how22}. 
\par
\begin{figure}

\centering
   \includegraphics[width=12.0cm, angle=0]{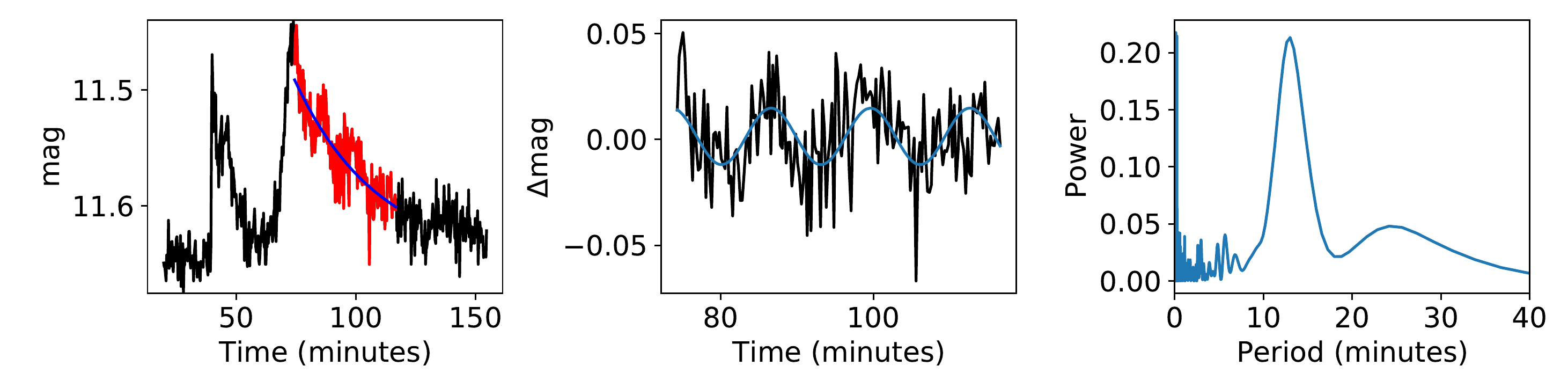}
\caption{The QPP of Star \#16 (TIC 29172363). Left panel: the QPP position in the flare is indicated in red. The blue curve is the 
fitted background. Middle panel: the QPP signal without background. The blue curve is the fitted QPP with a period of 13.0 
minutes; Right panel: the LS periodogram of the QPP signal in the middle panel.
\label{fig:qpp}}
\end{figure}

GWAC has a cadence of 15 seconds, shorter than TESS, and make it possible to find QPPs with short periods in GWAC white light flares. One flare occurring on Star \#16 (TIC 29172363) was found to indicate a QPP process as 
shown by the red light curve in the left panel of Figure \ref{fig:qpp}.  The function $f(t) = A_0 \times \exp(A_1t) + A_2$ was used to 
fit the background of the QPP signal, and shown by the blue curve in the left panel of Figure \ref{fig:qpp}. Here, $A_0, A_1$ and 
$A_2$ are parameters to be fitted, and $t$ is time in seconds. According to the light curve after subtracting the background 
information, a character manifested by QPP process is shown in the middle panel of Figure 5, and is analyzed by using the 
LS periodogram. The right panel of Figure \ref{fig:qpp} shows the periodogram result. A period of $13.0\pm1.5$ minutes is 
obtained for the QPP process.

\section{Binary and multiple systems} \label{sec:bin}
\begin{figure}

\centering
   \includegraphics[width=12.0cm, angle=0]{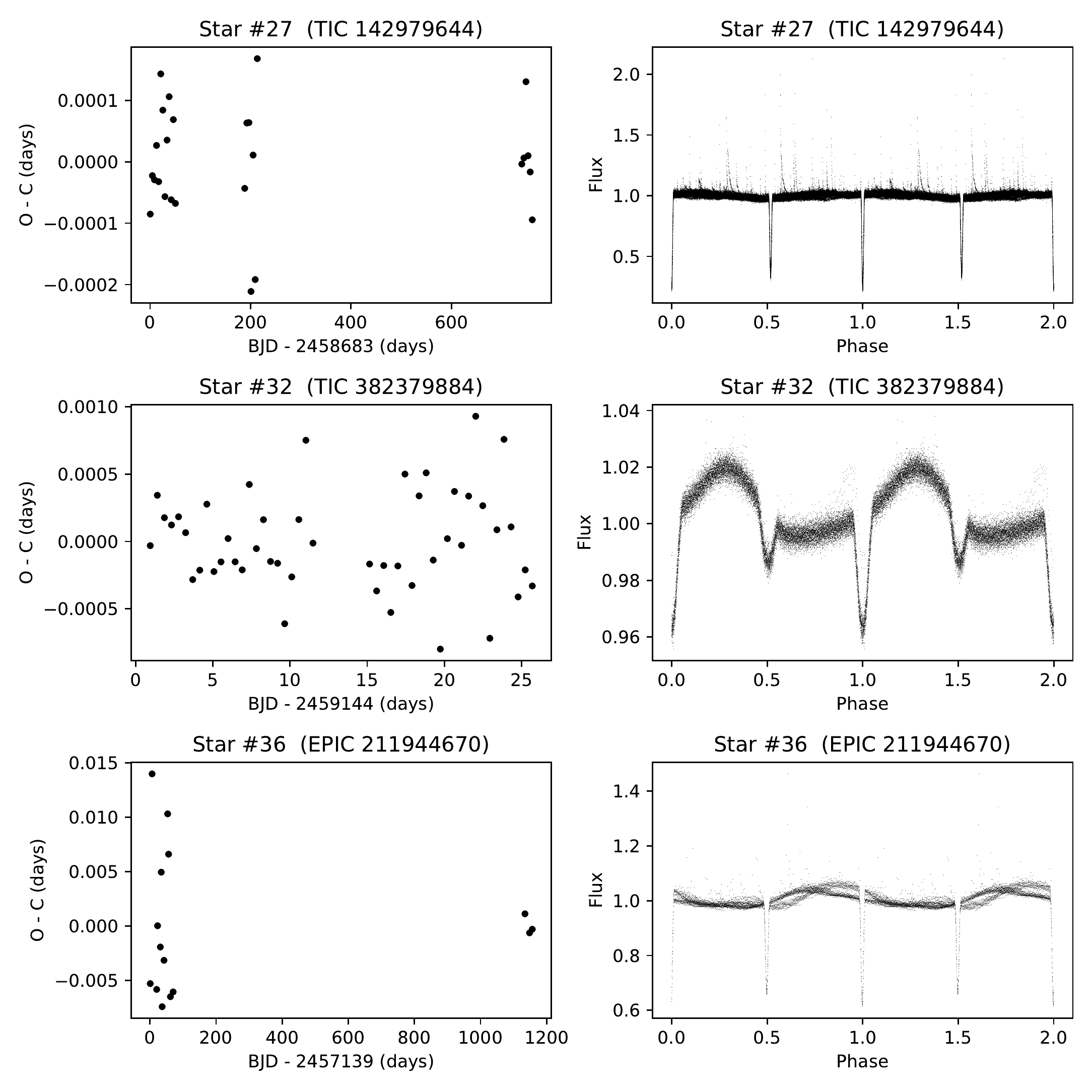}
\caption{O - C diagrams and light curves in phases of Star \#27 (TIC 142979644; 1RXS J120656.2+700754), \#32 (TIC 382379884) and \#36 (EPIC 211944670; CU Cnc) are shown in the left and right columns, respectively. 
\label{fig:oc}}
\end{figure}

We inspected all GWAC flare stars in Aladin \citep{bon00} and Gaia DR3, and found that there are 11 binaries and one triple 
system.  Among them, Star \#27 (TIC 142979644; 1RXS J120656.2+700754), \#32 (TIC 382379884) and \#36 (EPIC 211944670; CU Cnc) hold close eclipsing 
binaries. To obtain periods of binaries, we fitted the eclipse minimum times by quadratic polynomials of light curves, and 
ephemerides were calculated by fitting the eclipse minimum times. 
The ephemerides (in BJD) of Star \#27 (TIC 142979644; 1RXS J120656.2+700754), \#32 (TIC 382379884) and \#36 (EPIC 211944670; CU Cnc) are  
 $$T_{min} - 2457000 =  1683.36653104 (\pm 0.00002399) + 4.17903758 (\pm 0.00000027) E$$,
 $$T_{min} - 2457000 =  2144.48865796 (\pm 0.00010348) + 0.45825824 (\pm 0.00000316) E$$, and
 $$T_{min} - 2454833 =  2306.95408506 (\pm 0.00199451) + 2.77147142 (\pm 0.00001076) E$$, 
 respectively. Here, $E$ is the circle number, and $T_{min}$ is the eclipse minimum time. O - C diagrams and light curves in 
 phase are shown in Figure \ref{fig:oc}. Star \#36 (EPIC 211944670; CU Cnc) is a triple system in Gaia DR3, so its ephemerides may be disturbed by the 
 third star. 
 \par
 
The secondary eclipse minimum of Star \#27 (TIC 142979644; 1RXS J120656.2+700754) deviate from the phase 0.5 in Figure \ref{fig:oc}, so it has an eccentricity. We used the formulae 1 and 2 in \citet{lei22} to calculate its eccentricity ($e$) and 
periastron angle ($\omega$). From our calculation, its secondary eclipse phase $\phi_2 = 0.518$ (its primary eclipse phase 
$\phi_1 = 0$), and the widths of the primary and secondary eclipses were determined by eye, which are $w_1 \sim 0.0176$ and 
$w_2 \sim 0.018$, respectively. Then, $e \cos(\omega) = \frac{\pi}{2}\left[ \phi_2 -\phi_1 - 0.5 \right] = 0.028274$ and 
$e\sin(\omega) = \frac{w_2 - w_1}{w_2 + w_1} \sim 0.011236$. As a result, $e \sim 0.03$ and $\omega \sim 0.387$.
\par

\par
\begin{figure}
\centering
   \includegraphics[width=12.0cm, angle=0]{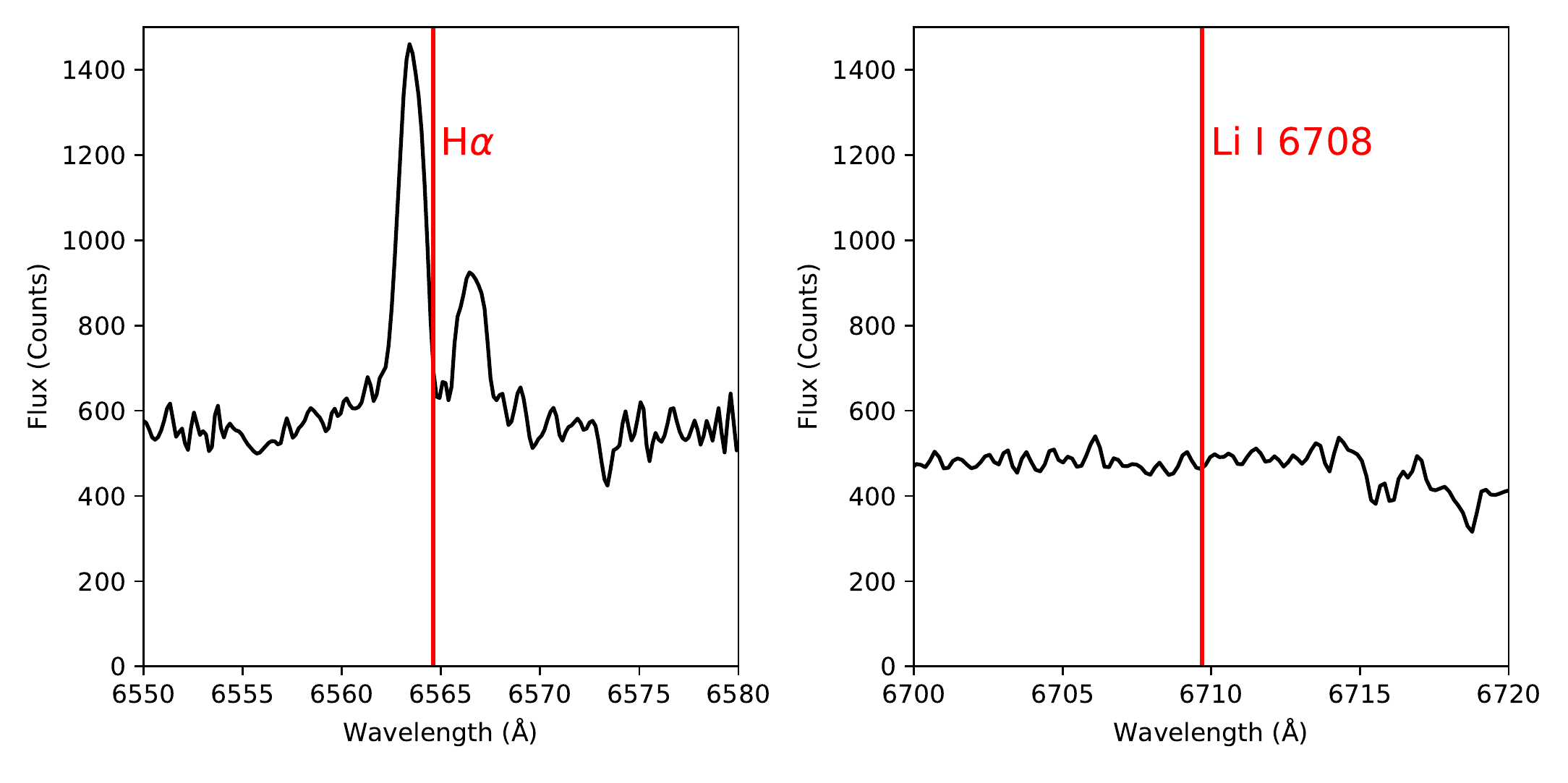}
\caption{The LAMOST MRS of  Star \#3 (HAT 178-02667). Left panel: Double H$\alpha$ emissions of Star \#3 (HAT 178-02667); Right panel: There is no Li I 6708. 
The positions of H$\alpha$ and Li I 6708 in vacuum are indicated in red lines.
\label{fig:s3}}
\end{figure}

Star \#3 (HAT 178-02667) has no Li I 6708 line in its LAMOST MRS as shown in the right panel of Figure \ref{fig:s3}, which implies that it is not a 
young star, and thus it unlikely holds a curcumstellar disk. Its $ruve = 7.65$ in Gaia DR3 and there are double H$\alpha$ 
emissions in its LAMOST MRS as shown in the left panel of Figure \ref{fig:s3}, so it is likely a binary system and double 
H$\alpha$ emissions indicate both components are active stars. In \citet{har11}, it has a period of 1.717885 days.

\section{H$\alpha$} \label{sec:ha}
\begin{figure}
\centering
   \includegraphics[width=12.0cm, angle=0]{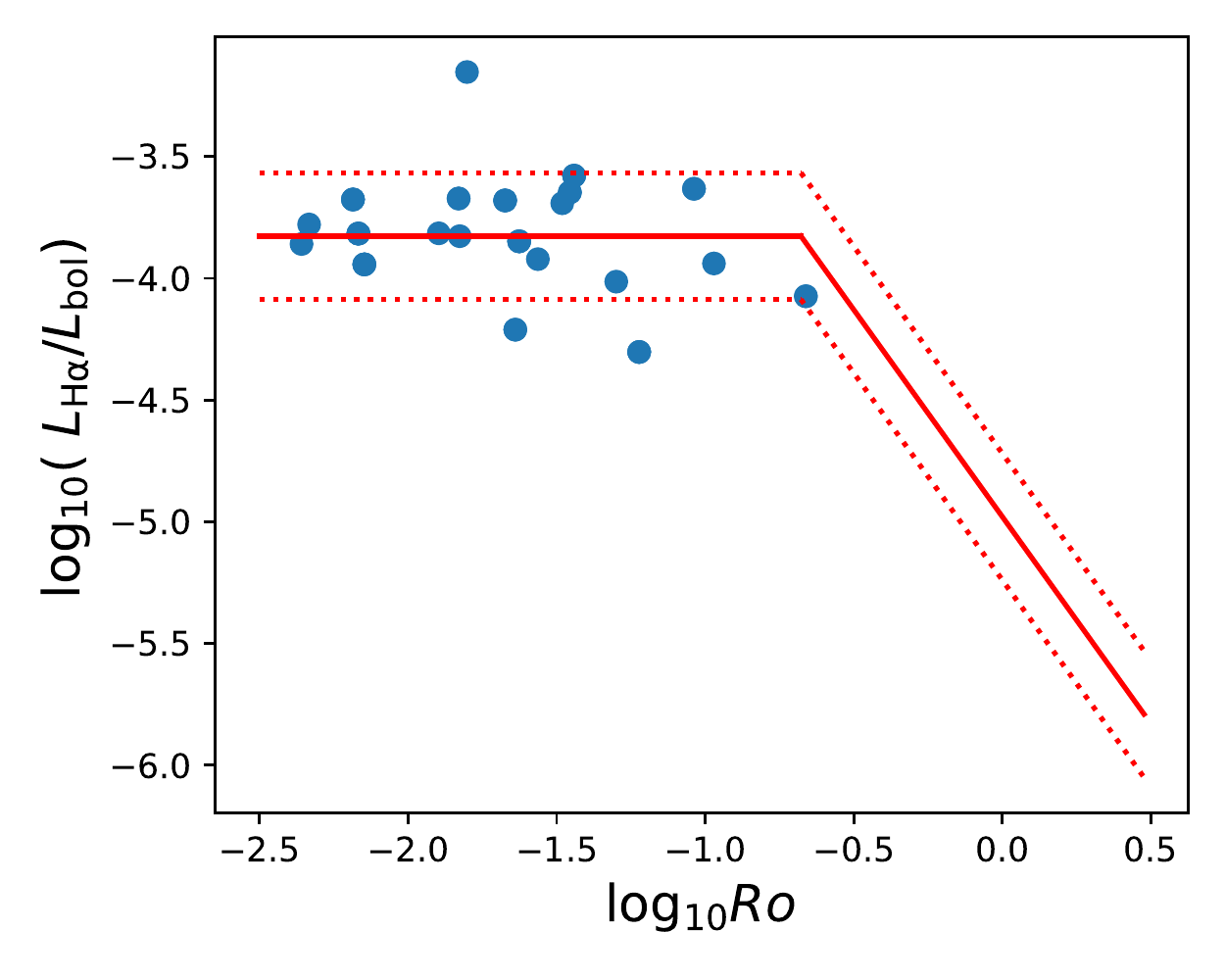}
\caption{The relationship between $Ro$ and $L_{\rm H\alpha}/L_{\rm bol}$. The red line and dotted lines are the relationship and 1$\sigma$ contours given in \citep{newton17}.
\label{fig:ha}}
\end{figure}

H$\alpha$ is a very important indicator of stellar activity. Stars with strong H$\alpha$ emissions in their quiescent spectra are very 
likely to show strong flare activities \citep{kow09}. Figure \ref{fig:ha} shows the relationship between H$\alpha$ luminosity from 44 
spectra of 21 M dwarfs and $Ro = P_{\rm rot}/\tau$, where $P_{\rm rot}$ is the stellar rotational period and $\tau$ is the 
convective turnover time calculated from Equation 10 in \citet{wri11}. For M type stars, their $L_{\rm X}/L_{\rm bol}$ shows 
saturation \cite[e.g.][]{wri18,piz03} and even supersaturation \cite[e.g.][]{jef11} for rapid rotators. Compared to Figure 7 in 
\citet{newton17}, our sample stars are all rapid rotators, and in the saturation region as shown In Figure 
\ref{fig:ha}.

\section{Habitablity} \label{sec:hab}
Stellar activity is a double-edged sword for life on a habitable planet. On one hand, flare activities can contribute the generation 
and development of life \citep{rim18}, on the other hand erode and even destroy the  ozone of a habitable-zone exoplanet 
\citep{til19}. We used Equation 10 in \citet{gun20} to delineate "abiogenesis zone," which means the flare frequency in this zone 
can contribute the prebiotic chemistry and then promote life generation. 
\par
The flare frequency distribution (FFD) is the cumulative flare energy frequency in per day \citep{ger72,gun20}, and often used to 
show how often a flare energy higher than a given value is. We obtained FFDs of 13 single stars with more than 20 flares 
detected in TESS or $K2$ light curves, and a power law function ${\rm log}_{10}(\nu) = \alpha {\rm log}_{10} (E_{\rm bol}) + 
\beta$ was used to fit each FFD, where $\nu$ is the cumulative flare frequency in day$^{-1}$, $E_{\rm bol}$ is the flare energy in 
erg, $\alpha$ and $\beta$ are parameters to be fitted. We also calculated the flare frequency $\nu$ of the GWAC flare energy 
using the fitted $\alpha$ and $\beta$ for each star, and found GWAC flares can occur at a frequency of 0.5 to 9.5\,yr$^{-1}$. 
\par
\begin{figure}
\centering
   \includegraphics[width=12.0cm, angle=0]{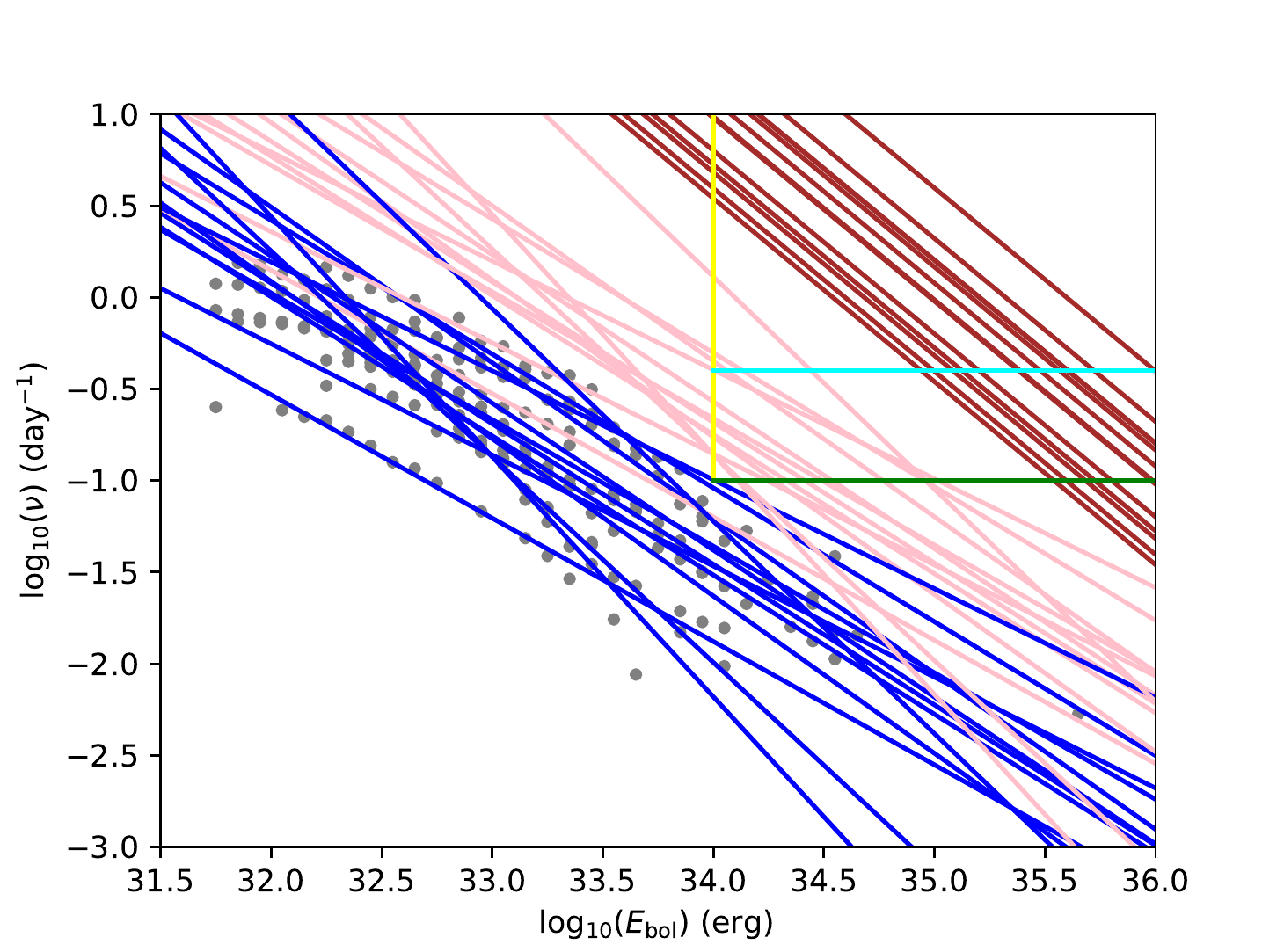}
\caption{FFDs of 13 flaring single stars with flare numbers greater than 20. The grey dots are the flare frequency obtained 
from TESS or $K2$ light curves, and their fitted power law functions are shown by blue lines for a flare temperature of 9000 K, 
while pink lines for 30000 K. The yellow line is log$_{10}(E_{\rm bol}) = 34$, and $\nu = 0.1$ and 0.4 are shown in cyan and green 
lines, respectively. The brown lines represent the abiogenesis zones for flare stars calculated by Equation 10 in \citet{gun20}.
 \label{fig:ffd}
}
\end{figure}

Flares with $E_{\rm bol} \geqslant 10^{34}$ erg can impact atmospheres of habitable planets. The cumulative flare frequency of 
$ \nu \geqslant 0.4$ day$^{-1}$ can remove more than 99.99\% of the ozone layer of a habitable-zone exoplanet  as  suggested 
by \citet{til19}, and a more permissive frequency limit is $ \nu \geqslant 0.1$ day$^{-1}$.Though the flare temperature of 9000 K is 
popularly used in literature, but the flare temperature can reach as high as 30000 K and even 42000 K \citep{how20b}. In Figure 
\ref{fig:ffd}, blue lines for a flare temperature of 9000 K, while pink lines for 30000 K. We can see that for the flare temperature of 
9000 K, two stars (TIC 88723334 and TIC 416538839) produce flares with energies greater than $10^{34}$ erg in the highest 
frequencies, but still can not destroy ozone layers of their habitable planets. However, for the flare temperature 30000 K, almost all 
stars can produce energetic flares to destroy ozone layers of their habitable planets, except Star \#28 (TIC 103691996). 

\par
Equation 10 in \citet{gun20} was also used to calculate the flare frequency limit for prebiotic chemistry. These frequency limits are 
shown by brown lines in Figure \ref{fig:ffd}, and there is no star having enough high energetic flares to trigger prebiotic chemistry 
on its habitable planets.

\section{Conclusion} \label{sec:con}
In this paper, we studied the 43 flares from 43 stars found in the GWAC archive between December 2018 and May 2019, by 
combining light curves from TESS and $K2$, spectra from LAMOST and the 2.16m telescope the Xinglong Observatory, and 
parallax and photometry from Gaia DR3, and obtained the following results:
\begin{enumerate}
\item We found 19 new active stars.
\item We found 3 sympathetic or homologous flares, which imply that the cool stars may share the similar physical magnetic 
explosions to those happening on the Sun.
\item We found a white light QPP in the sympathetic or homologous flare of Star \#16 (RX J0903.2+4207) with a period of $13.0 \pm 1.5$ minutes, 
which shows the advantage of GWAC with a cadence of 15 seconds in discovering white light QPPs with short periods.
\item Thirty-four stars have rotational or orbital  periods less than 5.4 days and only one star has a period of $\sim$10.42 days. 
\item Eleven stars are binaries and one is a triple system. The ephemerides of three binaries are calculated 
from their light curves, and one of them (Star \#27; 1RXS J120656.2+700754) also has an eccentricity of $e \sim 0.03$.
Star \#3 (HAT 178-02667) has no light curve, but double H$\alpha$ emissions in its LAMOST medium-resolution spectrum imply a binary. 
\item $L_{\rm H\alpha}/L_{\rm bol}$ shows that the rapid rotators in GWAC flare stars are in the  saturation region in the 
rotation-activity diagram.
\item Some of GWAC flare stars may produce enough energetic flares to destroy the ozone layer, but none can trigger prebiotic 
chemistry on its habitable planet.
\end{enumerate}
\par

Big flares with amplitudes greater than 1 mag detected by GWAC, can trigger telescopes in the Xinglong Observatory to follow 
up. Some research results have been obtained based on these observations \citep{xin21,wang21,wang22}. In future, we will 
continue to analysis these big flares to study their generation mechanisms and impacts on habitable planets.

\begin{landscape}

\tiny
 \begin{longtable}{lllllllllllllllll}
 \caption{GWAC flare stars}\label{tab:gwacflare}\\
\#& Name & Other Name&New&R.A. J2000 & Decl. J2000 &Dis& Gmag & bp-rp & SpT & Multi &Period &Date &Start&End&Flare Energy& ED  \\
&&&& degree & degree &pc & mag & mag & & &days &yyyymmdd&hhmmss&hhmmss& $\rm log_{10}(erg)$ &seconds\\ 
\hline\noalign{\smallskip}
1&TIC 141533801&PM J06521+7908&&103.040098&79.14838&39.98&10.514&1.68&M0&&5.20898(0.42182)&20190223&123107&141352&35.09(0.10)&4485  \\
2&TIC 8688061&PM J09108+3127&&137.701223&31.45744&27.04&12.506&2.653&M3&&4.24599(0.27568)&20190225&134702&150947&33.6(0.17)&2028  \\
3&&HAT 178-02667&Y&131.491539&36.59261&96.88&13.465&2.391&(M2.5)&Binary&1.717885&20190225&134702&150947&34.69(0.47)&4626  \\
4&TIC 253050844&G 176-59&Y&177.70409&45.56739&29.6&14.033&3.029&M4.5&&&20190403&144350&180450&33.82(0.59)&11285  \\
5&TIC 392402786&BD+48 1958&&174.352417&47.46249&33.47&10.034&1.687&M0&Binary&1.03001(0.01528)&20190403&144350&180450&34.98(0.09)&1851  \\
6&TIC 416538839&StKM 2-809&&183.914031&52.65242&25.1&11.373&2.608&M3&&0.72597(0.00680)&20190403&144350&180450&34.3(0.13)&4151  \\
7&TIC 334637014&LSPM J1542+6537&&235.554214&65.61809&38.92&13.233&2.889&M4.5&&0.61722(0.00399)&20190403&180612&205708&34.07(0.40)&5561  \\
8&TIC 162673744&HAT 149-01951&Y&246.061458&48.39066&59.69&13.615&2.482&M3&&1.63170(0.04786)&20190502&165833&200324&34.22(0.37)&4755  \\
9&TIC 436680588&2MASS J04542368+1709534&&73.598678&17.16485&143.62&16.617&2.365&(M2.5)&Binary&&20190113&102051&134136&35.37(0.51)&9090  \\
10&TIC 20161577&HAT 138-01877&Y&145.345992&43.97308&99.51&12.793&1.78&(M1)&&3.13022(0.14959)&20190125&151425&170400&34.19(0.08)&746  \\
11&TIC 88723334&G 196-3&Y&151.089429&50.38705&21.81&10.615&2.35&M2.5&&1.31341(0.02628)&20190125&151425&201525&34.56(0.09)&4946  \\
12&TIC 323688555&GJ 3537&&137.413115&6.70305&23&12.081&2.629&M3&&4.06548(0.30373)&20190224&135938&153823&33.36(0.10)&1068  \\
13&TIC 318230983&HAT 140-00487&Y&162.861825&48.69546&79.05&11.429&1.817&M1&&4.06913(0.25229)&20190124&175515&211330&35.53(0.19)&7433  \\
14&&1RXS J075908.2+171957&&119.779882&17.32982&24.16&12.706&2.8&(M4)&&0.48204(0.00411)&20190113&135650&185505&33.5(0.29)&2382  \\
15&EPIC 210701183&HAT 307-06930&Y&55.739582&18.37965&132.82&14.414&2.184&M1&&1.67409(0.01648)&20190112&102758&141043&34.75(0.69)&6870  \\
16&TIC 29172363&RX J0903.2+4207&&135.811479&42.12356&79.49&11.647&1.703&M0&&1.49055(0.03069)&20190201&170000&193443&35.44(0.13)&7350  \\
17&TIC 283729913&&Y&63.455228&9.21321&104.41&14.03&2.222&M2.5&&&20190114&113000&140052&34.36(0.42)&3188  \\
18&TIC 436614005&DR Tau&&71.775897&16.97856&192.97&11.651&1.621&T Tau&&&20190114&101252&154952&36.71(0.24)&23321  \\
19&TIC 274086357&1RXS J031750.1+010549&Y&49.457729&1.10221&151.67&12.219&1.38&K3&&0.63653(0.01102)&20181223&125610&162440&35.72(0.29)&6484  \\
20&TIC 427020004&HAT 307-01221&Y&49.572707&18.40566&94.92&11.693&1.495&K5&&3.16524(0.22178)&20190111&120004&151035&35.22(0.09)&3200  \\
21&TIC 114059158&&Y&55.304975&20.85436&140.32&13.862&1.96&M3&&0.31366(0.00193)&20181230&145636&162551&34.88(0.28)&4953  \\
22&TIC 195188536&DF Cnc&&128.870322&18.20561&49.23&12.906&2.376&M3&&5.35280(0.62707)&20190112&144958&195145&34.11(0.15)&2810  \\
23&TIC 77644831&1RXS J101627.8-005127&&154.112263&-0.86091&60.15&12.329&2.136&(M1.5)&&0.74709(0.00835)&20190114&155759&191159&34.58(0.11)&3307  \\
24&&&Y&153.304236&2.80498&105.46&17.593&3.115&(M5)&&&20190213&170041&180211&34.04(0.27)&1780  \\
25&TIC 374270454&1RXS J103715.4+020612&Y&159.314334&2.09753&71.84&11.829&1.951&M1&&3.51123(0.28641)&20190106&175650&205650&34.59(0.06)&1491  \\
26&TIC 142877499&G 236-81&&176.772653&70.03285&30.62&13.093&2.856&(M4)&Binary&10.42161(0.9949)&20190121&153702&205732&34.73(0.54)&16582  \\
27&TIC 142979644&1RXS J120656.2+700754&&181.73505&70.13054&17.37&10.892&2.845&(M4)&Binary&4.17903758(0.00000027)&20190318&134018&190333&33.65(0.12)&1065  \\
28&TIC 103691996&G 235-65&&154.787606&66.49275&29.26&13.053&2.662&(M3.5)&&&20190216&164856&195511&34.14(0.37)&9772  \\
29&TIC 99173696&&Y&123.681926&46.84348&39.3&13.745&2.84&M4&&1.78860(0.01522)&20190206&162410&185425&33.37(0.25)&1733  \\
30&TIC 153858162&1RXS J082204.1+744012&&125.533783&74.67285&47.66&12.264&2.266&(M3)&&1.97116(0.05626)&20190121&154417&183817&34.00(0.24)&1297  \\
31&TIC 270478293&LP 589-69&Y&34.197584&1.2126&31.51&12.476&2.532&M3&&4.49044(0.33040)&20181224&120134&142234&33.31(0.21)&735  \\
32&TIC 382379884&&Y&45.066475&-3.04574&114.87&12.073&1.53&M0&Binary&0.45825824(0.00000316)&20190127&102150&140010&35.48(0.14)&4360  \\
33&TIC 435308532&LP 413-19&&54.392052&17.85&38.8&12.079&2.712&M3&Binary&0.47630(0.00278)&20190204&113935&131650&34.12(0.06)&890  \\
34&TIC 440686488&V660 Tau&&57.116786&23.30076&134.02&12.242&1.319&K5&&0.23520(0.00104)&20190204&113935&131650&35.54(0.24)&5538  \\
35&TIC 457100137&1RXS J075554.8+685514&&118.972289&68.9069&29.2&13.056&2.844&(M4)&&0.53285(0.00445)&20&&&33.18(0.16)&1077  \\
36&EPIC 211944670&CU Cnc&&127.906559&19.39428&16.65&10.576&2.861&M3.5&Triple&2.7714842(0.00001076)&20190101&145240&190810&33.96(0.09)&1565  \\
37&TIC 224304406&1RXS J123415.2+481306&&188.564232&48.21862&46.83&13.135&2.675&M3&&0.94869(0.01295)&20190206&191205&221635&33.71(0.18)&1535  \\
38&&BX Ari&&44.546799&20.50087&234.85&11.921&1.541&K3&Binary&2.83690(0.14557)&20181229&135121&163451&36.35(0.16)&6342  \\
39&TIC 289040091&1RXS J064358.4+704222&&100.994199&70.70326&59.56&13.348&2.402&M3&&0.54374(0.00562)&20181215&162100&174019&34.42(0.62)&5981  \\
40&TIC 16246712&&Y&153.803156&37.86495&95.42&12.945&2.066&M1&Binary&&20&&&34.78(0.14)&1756  \\
41&TIC 445830121&&Y&173.045411&52.09011&42.3&13.465&2.503&M3&Binary&&20190106&210011&222626&33.59(0.48)&1790  \\
42&TIC 197251248&G 9-38&&134.558692&19.76258&5.15&11.966&3.777&M7&Binary&0.25397(0.00150)&20190211&150414&155014&32.51(0.15)&1704  \\
43&TIC 316276917&&Y&133.388088&56.78993&225.65&12.087&1.042&G7&&0.65862(0.00688)&20181227&165728&211943&35.56(0.15)&1797  \\
\noalign{\smallskip}\hline
\end{longtable}

\tablecomments{0.86\textwidth}{
1, The bolometric flare energies were calculated  from GWAC flares assuming a blackbody with a temperature of 9000 K. The fraction of the bolometric energy in the Gaia $G$ band is $\frac{f_G}{f_{\rm bol} } \approx 0.3$. \\
2, Spectral types in parentheses are assigned based on $G_{\rm rp} - G_{\rm bp}$. \\
3, The period of Star \#3 is from \citet{har11}. \\
}
\end{landscape}

\begin{table}
\bc
\begin{minipage}[]{100mm}
\caption[]{Data of which TESS Sector and K2 Campaign was used  \label{tab:sector}}\end{minipage}
\setlength{\tabcolsep}{2.5pt}
\small
 \begin{tabular}{lll}
  \hline\noalign{\smallskip}
Star \# & Name & Sectors \\  
1&TIC 141533801&19,20,26,40  \\
2&TIC 8688061&21  \\
3&&  \\
4&TIC 253050844&22  \\
5&TIC 392402786&22  \\
6&TIC 416538839&22  \\
7&TIC 334637014&14,15,16,17,21,22,23,24,41  \\
8&TIC 162673744&23,24,25  \\
9&TIC 436680588&(43),(44)  \\
10&TIC 20161577&21  \\
11&TIC 88723334&21  \\
12&TIC 323688555&8,34,45  \\
13&TIC 318230983&21  \\
14&&44,45,46  \\
15&EPIC 210701183&4  \\
16&TIC 29172363&21  \\
17&TIC 283729913&5,23  \\
18&TIC 436614005&(43),(44)  \\
19&TIC 274086357&4,31  \\
20&TIC 427020004&42,43  \\
21&TIC 114059158&42,43,44  \\
22&TIC 195188536&44,45  \\
23&TIC 77644831&8,35,45  \\
24&&  \\
25&TIC 374270454&35,45  \\
26&TIC 142877499&21  \\
27&TIC 142979644&14,15,21,41  \\
28&TIC 103691996&14,21,40,41  \\
29&TIC 99173696&20  \\
30&TIC 153858162&20,26,40,47,53  \\
31&TIC 270478293&4,31  \\
32&TIC 382379884&(4),31  \\
33&TIC 435308532&42,43,44  \\
34&TIC 440686488&42,43,44  \\
35&TIC 457100137&20,26,40  \\
36&EPIC 211944670&5,18  \\
37&TIC 224304406&22  \\
38& &(42),(43),(44)  \\
39&TIC 289040091&19,20,26  \\
40&TIC 16246712&21  \\
41&TIC 445830121&21,(22)  \\
42&TIC 197251248&44,45  \\
43&TIC 316276917&20  \\
  \noalign{\smallskip}\hline
\end{tabular}
\ec
\tablecomments{0.86\textwidth}{ The TESS sectors in parentheses mean that the data of these sectors are not available for this work.}
\end{table}

\begin{table}
\bc
\begin{minipage}[]{100mm}
\caption[]{H$\alpha$ luminosity. \label{tab:halum}}\end{minipage}
\setlength{\tabcolsep}{2.5pt}
\small
 \begin{tabular}{llll}
  \hline\noalign{\smallskip}
Star \# & R. A. J2000 & Decl. J2000 & $\lg (L_{\rm H\alpha}/L)$ \\
 & degree & degree &   \\
2&137.702323&31.457447&-3.78(+0.01),-3.79(+0.01),-3.58(+0.00) \\
4&177.704090&45.567390&-3.63(+0.01)\\
5&174.352814&47.462439&-5.05(+0.01)\\
6&183.914032&52.652434&-3.69(+0.00),-3.68(+0.00),-3.78(+0.01)\\
7&235.554210&65.618100&-3.15(+0.00)\\
8&246.061805&48.390525&-3.66(+0.01)\\
9&73.598465&17.162068&-4.22(+0.02)\\
11&151.089442&50.387056&-3.67(+0.01),-3.65(+0.00)\\
12&137.413190&6.703040&-3.63(+0.00),-3.64(+0.00),-3.68(+0.00)\\
13&162.861881&48.695440&-4.01(+0.01),-4.09(+0.01)\\
15&55.739577&18.379597&-3.83(+0.01),-3.92(+0.01),-3.94(+0.01)\\
16&135.811466&42.123555&-3.82(+0.01)\\
17&63.455420&9.213330&-4.30(+0.01)\\
18&71.775872&16.978558&-2.47(+0.01),-2.75(+0.01)\\
19&49.457706&1.102194&-3.80(+0.01)\\
20&49.572717&18.405648&-4.07(+0.02)\\
21&55.306192&20.854802&-3.48(+0.00),-4.60(+0.05),-3.56(+0.01),-3.83(+0.02)\\
22&128.870334&18.205612&-3.84(+0.01),-3.86(+0.01),-3.87(+0.01)\\
26&159.314354&2.097543&-4.21(+0.03),-4.21(+0.02)\\
30&123.681884&46.843252&-3.67(+0.00)\\
32&34.197623&1.212631&-3.94(+0.01)\\
35&57.116787&23.300774&-3.85(+0.01),-3.82(+0.01)\\
37&127.906510&19.394273&-3.78(+0.00),-3.76(+0.00),-3.73(+0.00)\\
38&188.564219&48.218640&-3.74(+0.01),-3.74(+0.01),-3.59(+0.01)\\
39&44.546799&20.500874&-3.58(+0.01)\\
40&100.993690&70.702840&-3.69(+0.00),-3.82(+0.01)\\
41&153.802305&37.864154&-3.76(+0.02),-3.98(+0.05),-3.61(+0.02)\\
  \noalign{\smallskip}\hline
\end{tabular}
\ec
\tablecomments{0.86\textwidth}{Each spectrum provides one $\lg (L_{\rm H\alpha}/L)$, with the error in the following parenthesis. }
\end{table}

\normalem
\begin{acknowledgements}
We thank our anonymous referee for the insightful comments.
This work is supported by the National Natural Science Foundation of China (NSFC) with grant No. 12073038.
This work is supported by the Joint Research Fund in Astronomy U1931133 under cooperative agreement between the National Natural Science Foundation of China (NSFC) and Chinese Academy of Sciences (CAS).
Chen Yang, Chao-Hong Ma, Xu-Kang Zhang, Xin-Li Hao, and Xiao-Feng Meng acknowledges the NSFC with grant No. 61941121. 
Jie Chen acknowledges the Beijing Natural Science Foundation, No. 1222029.
\par
Guang-Wei Li thanks Dr. Ting Li for helpful discussions.
\par
Guoshoujing Telescope (the Large Sky Area Multi-Object Fiber Spectroscopic Telescope LAMOST) is a National Major Scientific Project built by the Chinese Academy of Sciences, Funding for the project has been provided by the National Development and Re- form Commission. LAMOST is operated and managed by the National Astronomical Observatories, the Chinese Academy of Sciences.
\par
We acknowledge the support of the staff of the Xinglong 2.16m telescope. This work was partially supported by the Open Project Program of the Key Laboratory of Optical Astronomy, National Astronomical Observatories, Chinese Academy of Sciences. 
\par
This research has made use of "Aladin sky atlas" developed at CDS, Strasbourg Observatory, France.

\end{acknowledgements}
  
\bibliographystyle{raa}
\bibliography{rdflareraa}

\end{document}